**ORIGINAL ARTICLE** OPEN ACCESS

# The Future of AI in the GCC Post-NPM Landscape: A Comparative Analysis of Kuwait and the UAE

Mohammad Rashed Albous[1] | Bedour Alboloushi[2] | Arnaud Lacheret[3]

[1]Abdullah Al Salem University, Khalidiya, Kuwait | [2]Kuwait College of Science and Technology, Doha, Kuwait | [3]SKEMA Business School, Université Côte d'Azur, Sophia Antipolis, France

**Correspondence:** Mohammad Rashed Albous (mohammad.albous@aasu.edu.kw)

## ABSTRACT

Comparative evidence of how two Gulf Cooperation Council (GCC) states translate artificial intelligence (AI) ambitions into post–New Public Management (post-NPM) outcomes are scarce because most studies focus on Western democracies. To fill this gap, we examine constitutional, collective choice, and operational rules that shape AI uptake in two contrasting GCC members, the United Arab Emirates (UAE) and Kuwait, and whether they foster citizen centricity, collaborative governance, and public value creation. Anchored in Ostrom's Institutional Analysis and Development framework, the study integrates a most similar/most different systems design with multiple sources: 62 public documents issued between 2018 and 2025, embedded UAE cases (Smart Dubai and MBZUAI), and 39 interviews with officials conducted from Aug 2024 to May 2025. Dual coding and process tracing connect rule configurations to AI performance. Our cross-case analysis identifies four mutually reinforcing mechanisms behind divergent trajectories. In the UAE, concentrated authority, credible sanctions, pro-innovation narratives, and flexible reinvestment rules transform pilots into hundreds of operating services and significant recycled savings. Kuwait's dispersed veto points, exhortative sanctions, cautious discourse, and lapsed AI budgets, by contrast, confine initiatives to pilot mode despite equivalent fiscal resources. These findings refine institutional theory by showing that vertical rule coherence, not wealth, determines AI's public value yield, and temper post-NPM optimism by revealing that efficiency metrics advance societal goals only when backed by enforceable safeguards. To curb ethics washing and test the transferability of these mechanisms beyond the GCC, future research should track rule diffusion over time, experiment with blended legitimacy-efficiency scorecards, and investigate how narrative framing shapes citizen consent for data sharing.

**Related Articles:**

Robles, P. and D. J. Mallinson 2023. "Catching up With AI: Pushing Toward a Cohesive Governance Framework." *Politics & Policy*, 51, no. 3: 355–372. https://doi.org/10.1111/polp.12529.

Veloso Meireles, A. 2024. "Digital Rights in Perspective: The Evolution of the Debate in the Internet Governance Forum." *Politics & Policy* 52, no. 1: 12–32. https://doi.org/10.1111/polp.12571.

Zeng, J., T. Stevens, and Y. Chen. 2017. "China's Solution to Global Cyber Governance: Unpacking the Domestic Discourse of 'Internet Sovereignty.'" *Politics & Policy* 45 no. 3: 432–64. https://doi.org/10.1111/polp.12202.

**摘要**

由于大多数研究侧重于西方民主国家，关于海湾合作委员会(GCC)两个成员国如何将人工智能(AI)雄心转化为后新公共管理(post-NPM)成果的比较证据稀缺。为了填补这一空白，我们考察了宪法、集体选择和运营规则如何影响两个截然不同的海湾合作委员会成员国——阿拉伯联合酋长国(UAE)和科威特——对人工智能的接受程度，以及它们是否促进了以公民为中心、协作治理和公共价值创造。本研究以奥斯特罗姆的"制度分析与发展"框架为基础，整合了"最相似/最不同"的系统设计，并包含多个来源：2018年至2025年间发布的62份公共文件、阿联酋案例(SmartDubai和MBZUAI)以及2024年8月至2025年5月进行的39次官员访谈。双重编码和过程追踪将规则配置与人工智能绩效联系起来。我们的跨案例分析确定了不同发展轨迹背后四种相互强化的机制。阿联酋凭借集中的权力、可信的制裁措施、有利于创新的叙事和灵活的再投资规则，将试点项目转化为数百项运营服务，并带来可观的循环节约。相比之下，科威特分散的否决点、劝诫式的制裁、谨慎的论述以及人工智能预算的失效，尽管财政资源相当，却将相关举措限制在试点模式。这些研究结果完善了制度理论，表明垂直规则的一致性（而非财富）决定了人工智能的公共价值收益；这些研究也缓和了后新公共管理时代的乐观情绪，表明只有在可执行的保障措施支持下，效率指标才能促进社会目标的实现。为了遏制伦理洗白，并测试这些机制在海湾合作委员(GCC)以外的可迁移性，未来研究应追踪规则随时间推移的扩散情况，尝试混合合法性-效率记分卡，并探究叙事框架如何影响公民对数据共享的同意。

**RESUMEN**

La evidencia comparativa de cómo dos estados del Consejo de Cooperación del Golfo (CCG) traducen sus ambiciones en materia de inteligencia artificial (IA) en resultados posteriores a la Nueva Gestión Pública (post-NPM) es escasa, ya que la mayoría de los estudios se centran en las democracias occidentales. Para subsanar esta deficiencia, examinamos cómo las normas constitucionales, de elección colectiva y operativas influyen en la adopción de la IA en dos miembros del CCG con características dispares, los Emiratos Árabes Unidos (EAU) y Kuwait, y si estas fomentan la centralidad ciudadana, la gobernanza colaborativa y la creación de valor público. Basado en el marco de Análisis y Desarrollo Institucional de Ostrom, el estudio integra un diseño de sistemas de máxima similitud/máxima diferencia con múltiples fuentes: 62 documentos públicos emitidos entre 2018 y 2025, casos integrados de los EAU (SmartDubai y MBZUAI) y 39 entrevistas con funcionarios realizadas entre agosto de 2024 y mayo de 2025. La codificación dual y el rastreo de procesos vinculan las configuraciones de las normas con el rendimiento de la IA. Nuestro análisis de casos cruzados identifica cuatro mecanismos que se refuerzan mutuamente tras las trayectorias divergentes. Los Emiratos Árabes Unidos, con una autoridad concentrada, sanciones creíbles, narrativas proinnovación y normas flexibles de reinversión, transforman los proyectos piloto en cientos de servicios operativos y un importante ahorro reciclado. En contraste, la dispersión de los puntos de veto, las sanciones exhortativas, el discurso cauteloso y los presupuestos de IA caducados de Kuwait limitan las iniciativas a la fase piloto a pesar de contar con recursos fiscales equivalentes. Estos hallazgos refinan la teoría institucional al demostrar que la coherencia vertical de las regras, y no la riqueza, determina el rendimiento del valor público de la IA, y moderan el optimismo posterior a la NGP al revelar que las métricas de eficiencia solo promueven los objetivos sociales cuando están respaldadas por salvaguardas exigibles. Para frenar el lavado de imagen ética y comprobar la transferibilidad de estos mecanismos más allá del CCG, las investigaciones futuras deberían rastrear la difusión de las reglas a lo largo del tiempo, experimentar con cuadros de mando combinados de legitimidad y eficiencia e investigar cómo el encuadre narrativo configura el consentimiento ciudadano para el intercambio de datos.

## 1 | Introduction

For more than 30 years, public administrators have grappled with reconciling evolving technological frontiers, particularly in artificial intelligence (AI), with broader shifts in governance philosophy. Early reformers championed New Public Management (NPM) for its market-like mechanisms and performance metrics (Christensen and Lægreid 2022; Hood 1991; Osborne and Gaebler 1992). Critics later observed that these external rules often neglected the patterns of reciprocity, equity, and co-production that emerge when citizens and agencies interact repeatedly (Denhardt and Denhardt 2015; Ongaro 2024; Treiber 2023). In response, post-New Public Management (post-NPM) reforms have emerged, with scholars highlighting three core goals that are especially relevant to our discussion: citizen-centricity, collaborative governance, and public value creation (Ansell and Torfing 2021; Moore 1997; Pollitt and Bouckaert 2017). These objectives view citizens as co-producers, encourage cross-sector partnerships, and elevate benefits that resonate throughout a community rather than within a single bureau.

AI technologies, from large language models (LLMs) that sift text at scale to predictive analytics that allocate scarce resources, now offer public managers new tools for pursuing those goals (Mergel et al. 2024; OECD 2024). Routing algorithms can lower the cost of equitable service delivery, integrated data platforms can knit together fragmented agencies, and forecasting models can anticipate demand before shortages appear (Chen 2024; Neumann et al. 2024). Recent cross-jurisdiction reviews show that algorithmic-accountability regimes, such as the binding EU AI Act, remain uneven in scope and enforcement when contrasted with the largely voluntary provisions found in GCC National AI Strategy (NASs) (Lazar 2025; Schultz et al. 2024). Yet, the empirical record remains anchored in Western democracies, leaving important knowledge gaps about how AI is taken up where rule configurations, social norms, and political incentives differ (Roche et al. 2023).

That gap is especially salient in the Gulf Cooperation Council (GCC), a region endowed with fiscal resources yet characterized by institutional diversity. Formal statutes often coexist with strong tribal affiliations, and data privacy norms evolve alongside ambitious digital-economy plans (Al-Kuwari 2025; Almutairi 2024). The result is a rich laboratory for examining how formal rules, informal norms, and shared resource pools condition collective action around AI in the GCC region (Ateeq et al. 2025). Within this context, we compare Kuwait and the United Arab Emirates (UAE) because they hold fiscal capacity, broadband penetration, and population size roughly constant while differing sharply in their rule architectures, giving us leverage to isolate the institutional drivers of AI uptake.

This study therefore undertakes a comparative inquiry into two GCC member-states: the UAE and Kuwait. The UAE illustrates a top-down, centrally orchestrated rule system backed by the UAE NAS 2031, unveiled in 2018, under the direct stewardship of the Minister of State for AI. Kuwait, by contrast, issued its first NAS only in 2024 and has proceeded more cautiously, shaped by parliamentary scrutiny and widespread concern for data privacy and trust (Albous et al. 2025; Hendawy and Kumar 2024; Sarker and Rahman 2023). Much as differing irrigation bylaws yield contrasting resource outcomes across neighboring villages, these two designs supply a natural experiment for observing how AI reforms fare under alternative rule sets.

Following Ostrom's IAD framework, constitutional rules assign decision-making authority, collective-choice rules specify how that authority is exercised, and operational rules govern daily implementation. This study asks: How do these rule types shape AI adoption in Kuwait and the UAE, and in turn advance the post-NPM goals of citizen-centricity, collaborative governance, and public value creation? By answering this question, we aim to deepen both theory and practice by showing how institutional configurations and social norms in resource-rich, non-Western settings condition the public value that AI can deliver.

Institutional theory (North 1990; Scott 2013) provides the overarching proposition that durable rule systems shape organizational behavior. We translate that proposition into concrete analytical categories via Ostrom's Institutional Analysis and Development (IAD) framework, a widely used rule-mapping extension of institutional theory (Imperial 1999). The framework's constitutional, collective choice, and operational rule tiers supply our coding lens (Morgan et al. 2023; Schlager and Villamayor-Tomas 2023). Consistent with Ostrom's emphasis on rules-in-use and nested arenas (Ostrom 2011; Lewis and Aligica 2024), we trace how legislation, monitoring arrangements, and shared norms mediate technology uptake, an approach recently demonstrated in Yuan and Chen's (2025) analysis of algorithmic accountability. Resource abundance and executive sponsorship may launch pilot projects, yet their survival hinges on the wider institutional fabric that assigns authority, resolves disputes, and fosters trust.

Empirically, we rely on a qualitative, multi-method design that weaves together document analysis, two embedded case studies, and thirty-nine semi-structured interviews with Emirates and Kuwaiti officials. By tracing how AI initiatives interact with existing rule systems, we identify mechanisms that either unlock or constrain their capacity to produce citizen-oriented, collaborative, value-creating outcomes.

This study advances the AI governance and post-NPM literatures in three ways. First, it operationalizes Ostrom's IAD as a rule-mechanism map for AI governance, tracing constitutional, collective-choice, and operational rules to outcomes and surfacing four mechanism bundles: rule concentration, sanction credibility, narrative alignment, and resource discipline. Second, it offers a hybrid test by tracking classic NPM efficiency indicators (time-on-task saved, cost per transaction, outsourcing intensity) alongside post-NPM outcomes (citizen-centricity, collaborative governance, public value), showing when these logics complement or substitute. Third, it distills design lessons that translate the heuristic "rules beat resources" into actionable reforms for rule coherence, enforceable safeguards, and reinvestment, with concrete guidance for centralized systems seeking stronger oversight.

We focus on policy processes and outcomes (how constitutional, collective choice, and operational rules shape AI adoption and public value) rather than technical model performance.

The study proceeds as follows. Section 2 reviews scholarship on AI in public administration, the post-NPM turn, Gulf institutional diversity, and Ostrom's IAD framework, and distills these strands into the analytical model that guides the inquiry. Section 3 sets out our most similar/most different comparative logic, multi-stream data collection, coding pipeline, and the validity and ethical safeguards built into the research design. Section 4 reports the empirical results in four blocks: document corpus, embedded cases, interview evidence, and cross-case process tracing, showing how rule configurations in the UAE and Kuwait shape AI's contribution to citizen centricity, collaborative governance, and public value creation. Section 5 interprets those findings for the hybrid NPM/post-NPM debate, articulates the "rules beat resources" proposition, and extracts design lessons for governments that must balance speed, legitimacy, and accountability. Section 6 concludes by taking stock of the study's scope conditions, synthesizing its theoretical contributions and practical governance takeaways, and mapping a research agenda that links institutional coherence, narrative framing, and AI talent flows.

## 2 | Literature Review

Scholarship on digital government in the Gulf has expanded rapidly, yet much of it treats each state in isolation or assumes a single "GCC model." To provide a firmer comparative yardstick, the present review is organized in six tightly linked steps. First, it traces the region's journey from early e-government portals to today's smart-city systems, foregrounding the contrasting rule configurations that guide the UAE and Kuwait. Second, it weighs the collective gains and collective risks that AI introduces into public administration, with particular attention to data-protection gaps and "ethics-washing." Third, it revisits post-NPM goals, citizen-centricity, collaborative governance, and public value creation through a collective-action lens that highlights the incentives and constraints policymakers face. Fourth, it outlines an institutional framework that distinguishes constitutional, collective-choice, and operational rules, setting the analytical foundations for comparison. Fifth, it specifies the concrete indicators, anchored in Moore's Strategic Triangle, that translate those goals into observable measures. Sixth, it positions the study within the wider literature by explaining why a side-by-side examination of two rule systems can advance both comparative digital-government research and the evolving post-NPM debate.

## 2.1 | From E-Government Islands to Smart-City Commons in the GCC

GCC governments began digitizing public services in the early 2000s, launching one-stop portals that lowered transaction costs for routine tasks such as vehicle registration (Al Ali et al. 2023; Al-Hajri et al. 2024). Many portals, however, functioned as stand-alone "islands of automation," lacking shared data vocabularies, payment rails, or grievance channels (Hale et al. 1989; Razavi and Habibnia 2024). A second wave, branded smart governance, sought to stitch those islands together (Dickinson and Yates 2023; Madan and Ashok 2023). Dubai's Smart Dubai program, created by royal decree, mandates agency participation in a central data layer and assigns a single office to monitor compliance (Digital Dubai 2020). Kuwait's Sahel platform launched in 2021 follows a more polycentric path: each ministry retains veto power over data sharing, and parliamentary committees periodically revise privacy provisions, a pattern that induces regulatory caution and slows government willingness to integrate data across agencies (Albous 2024; Alhosani and Alhashmi 2024).

Most regional surveys still merge the Gulf states into a single "GCC model" (Ateeq et al. 2025), obscuring important institutional variation. To reveal those differences, this study treats the UAE and Kuwait as separate rule systems. Aggregation conceals how variations in monitoring rights, dispute-resolution forums, and sanctioning tools shape platform performance. Western syntheses, by contrast, typically presuppose robust data-trust regimes such as the GDPR and the EU AI Act, both anchored in the European Charter of Fundamental Rights (Csernatoni 2024). Gulf governments rely instead on NASs, executive decrees, or other largely voluntary codes. Whether these contrasting legal traditions translate into measurably different citizen outcomes remains an open empirical question.

## 2.2 | AI in Public Administration: Collective Promise and Collective Risk

AI applications can help public agencies reallocate scarce resources, forecast service demand, and personalize interactions with citizens (Dixon et al. 2018; Mergel et al. 2024). In the UAE, the Happiness Meter aggregates user feedback from more than 4000 service touchpoints and directs managerial attention to under-performing bureaus (Digital Dubai 2023; El Khatib et al. 2023; ITU 2019). In Kuwait, Parliament and Cabinet have repeatedly debated the introduction of AI-enabled transportation systems, yet no pilot projects have been launched and no supporting procurement rules have been ratified (Hamwi et al. 2022; Sadriwala et al. 2024).

As a complementary capacity yardstick, we also reference the IFF Global AI Competitiveness Index (Forum on Information & Democracy 2024). For cross-country infrastructure readiness, we rely on the International Telecommunication Union (ITU) ICT Development Index (ITU 2023, 2024) as a comparative baseline. For AI talent-migration indicators, we draw on the Stanford AI Index 2025 (Maslej et al. 2025). For macro-level diversification context in the GCC, see the Global Economic Diversification Index (Prasad et al. 2024). As an innovation-capacity baseline, we also reference the Global Innovation Index (WIPO 2024).

International evidence warns that algorithms can reproduce historical inequities when their training data contain embedded biases (Green 2022) and that voluntary ethics codes often lapse into "ethics-washing" when enforcement incentives are weak (Schultz et al. 2024). Ethics-washing typically arises when non-binding instruments, most notably NASs that laud ethical principles but impose no sanctions, stand in for hard law or independent oversight. Fragmented regulations in the GCC amplify these risks. Coeckelbergh (2024) and Hagendorff (2020) draw attention to concerns that go beyond algorithmic bias and ethics-washing. These include practical and often neglected matters such as how AI systems might be misused for political abuse, the lack of diverse voices in the development of these technologies, and the wider social and environmental impacts they may have.

The UAE's Data Protection Law (PDPL) (2021) applies across the public sector, whereas Kuwait's Data Privacy Protection Regulation (DPPR) (2024) covers only telecom and IT providers, leaving other governmental and non-governmental sectors without clear legal guidance. Private actors echo these governance frictions: a 2025 Mohammed bin Rashid School of Government (MBRSG) survey conducted with Google found that nearly 70% of 81 UAE-based AI firms cite regulatory uncertainty as their main scaling barrier, despite abundant funding, illustrating how fast-shifting "soft law" can still deter innovation even in a top-down system (Salem and Shaer 2025). A distinct but related concern appears in Kuwait. High-profile lapses matter here: Amnesty International labeled Kuwait's 2020 COVID-19 tracing app "a highly invasive surveillance tool," and MPs still invoke that report when data-sharing bills reach committee hearings (Amnesty International 2020).

Cross-border data exchange within the GCC still relies on scattered bilateral memoranda rather than a unified statute (Al-Kuwari 2025; Alshehhi et al. 2024). Without region-wide rules for monitoring and sanctioning, incentives to share data remain weak, and the prospect of a genuinely collective AI ecosystem stays out of reach, even in a bloc whose raison d'être is cooperation, not competition. No comparative study has yet measured how this regulatory asymmetry affects AI deployment speed or citizen trust across Gulf states.

## 2.3 | Post-NPM Goals in a Hybrid Governance Landscape

Post-NPM reformers emphasize three overarching aims: citizen-centricity, collaborative governance, and public-value creation (Ansell and Torfing 2021). Digital platforms can advance these aims; yet, results are uneven. Large language chatbots shorten response times, but if users cannot appeal automated decisions, perceived fairness may decline (Zuiderwijk et al. 2021). Shared data lakes can knit agencies together; yet, only when contributors accept joint rules for cost-sharing, security, and audit (Molodtsov and Nikiforova 2024).

Importantly, post-NPM ideas have not displaced the core logics of NPM. Performance metrics, market analogies, and contracting still shape many reform programs (Dan et al. 2024; Dunleavy et al. 2006). Recent scholarship therefore depicts a hybridization in which NPM efficiency norms coexist with post-NPM participation norms (Bel and Casula 2024; Goldfinch and Halligan 2024). See also Funck and Karlsson (2024) on how "social imaginaries" of governance innovation mix these logics, and Funck (2025) on trust-based management controls that replace rigid performance audits. Some label this synthesis "New Public Governance," organized around legitimacy, trust, and capacity as its three pillars (Kann-Rasmussen 2023; Casady et al. 2019). Where these layers complement one another, tight performance targets plus robust citizen feedback, digital projects gain traction; where they collide, real-time dashboards versus lengthy parliamentary scrutiny, roll-outs stall or proceed selectively. Comparative Gulf evidence on such frictions remains limited, a gap the present inquiry addresses. Our study will privilege the hybridization approach.

While post-NPM priorities anchor this inquiry, the managerial doctrines of NPM have by no means disappeared from the GCC reform playbook. Empirical evidence already points to a hybrid governance layer in which classic efficiency logics (cost, speed, outsourcing) coexist with, and at times condition, the pursuit of post-NPM goals. To capture that residual layer, we track three NPM indicators across our embedded cases: (i) time on task saved (labor hours eliminated through automation), (ii) cost per transaction (average AED per completed e-service relative to the former paper channel), and (iii) outsourcing intensity (share of project spend or MoUs contracted to private/foreign vendors). These metrics allow us to test whether NPM tools reinforce or undermine citizen-centricity, collaborative governance, and public value creation.

## 2.4 | Institutional Rules-In-Use for AI Governance

Institutional theory offers the high-level proposition that durable rule configurations shape organizational behavior and policy outcomes (North 1990; Scott 2013). Both institutional theory and Ostrom's IAD framework recognize that institutions can persist or evolve through processes of interaction, feedback, and learning. To translate that proposition into a tractable coding scheme, we adopt Ostrom's IAD framework, which parses institutions into constitutional, collective choice, and operational rule tiers (Ostrom 2005; Ostrom et al. 1994). Decades of empirical work confirm that IAD serves as a widely accepted rule-mapping extension of institutional theory in domains as diverse as ecosystem management (Glynn and D'aunno 2023), Ostrom's framework has informed water governance (Pahl-Wostl et al. 2010), urban commons (Foster and Iaione 2015), open-data policy (Litschka and Pellegrini 2019), and, most recently, algorithmic accountability studies (Yuan and Chen 2025). By importing that rule-mapping tool into the study of AI in Gulf public administration, we can move from abstract notions of authority and legitimacy to observable design variables such as who may issue AI directives, how monitoring rights are allocated, and which sanctions backstop data-sharing agreements. Those variables, in turn, enable a systematic test of whether different rule configurations advance the post-NPM goals of citizen-centricity, collaborative governance, and public value creation that structure our inquiry.

Within institutional theory, Ostrom's IAD framework distinguishes three nested sets of rules: constitutional rules that allocate authority to craft policy, collective choice rules that specify how those policies are made, and operational rules that govern day-to-day implementation (Ostrom et al. 1994; Panzaru 2024; Thornton and Ocasio 2008). In the UAE, a constitutional rule assigns the Minister of State for AI authority to issue binding directives, while a single federal office monitors compliance. In Kuwait, constitutional authority is dispersed among the Cabinet, parliament, and multiple sector regulators, introducing veto points that slow adoption.

Recent GCC research confirms the value of this lens. Chandran et al. (2023) show that AI pilots flourish when professional bodies hold clear monitoring rights. Kuwait offers the opposite showcase: overlapping mandates between the Central Agency for Information Technology (CAIT) and the Communications and Information Technology Regulatory Authority (CITRA) create ambiguous collective choice rules, delaying inter-agency data exchange and stalling several AI proposals (Raji et al. 2022; Ruijer et al. 2023). Yet no comparative work has systematically mapped UAE and Kuwaiti rule structures against AI outcomes, nor traced how informal norms, tribal affiliation, privacy expectations, and leadership narratives reshape the formal statutes.

The analytical framework set out in Table 1 aligns each IAD tier with the three post-NPM objectives and details the concrete UAE and Kuwaiti instruments that shape AI uptake in the two states. This framework guides the empirical analysis that follows.

## 2.5 | Positioning the Study

Most empirical work on AI and public administration still comes from Western democracies, where oversight routines and privacy norms are relatively settled (Green 2022; Roche et al. 2023). A growing body of GCC scholarship is emerging, yet it remains sector-specific and rarely tackles cross-state comparisons. For example, Jarab et al. (2025) show that public perceptions of AI in UAE health care are shaped by concerns about data security, trust, and the humanistic aspects of care, but they do not examine how regulatory frameworks or rule systems influence these concerns across the region.

The GCC therefore offers a valuable natural laboratory: fiscal abundance coexists with divergent constitutional structures and social expectations. By setting the UAE's centrally orchestrated design alongside Kuwait's more polycentric, parliament-modulated arena, we ask how alternative rule configurations condition AI's contribution to citizen-centricity, collaborative governance, and public value. Applying Ostrom's institutional rule typology to these two cases moves the discussion beyond single-country portraits and tests whether similar technologies flourish or falter in different collective-action settings. In doing so, the study extends comparative digital-government research and sharpens the post-NPM debate for resource-rich, non-Western contexts.

TABLE 1 | Institutional rule matrix for AI governance in the UAE and Kuwait.

| Rule tier (IAD) | Contribution to post-NPM | UAE – rules/key instruments | Kuwait – rules/key instruments | Rationale/implication |
|---|---|---|---|---|
| **Constitutional** (who may decide) | **Citizen-centricity**. Authorize nationwide digital IDs and service portals that widen access<br><br>**Collaborative governance**. Give legal force to cross-agency data exchange and joint AI initiatives<br><br>**Public-value creation**. Signal top-level legitimacy and earmark long-term resources | • Federal Cabinet Decision (2017) creates Minister of State for AI and National Program for AI (NAS 2031)<br>• Federal Law (2019) establishes MBZUAI<br>• Cabinet Resolution (2025) launches AI Legislative Intelligence Office for "statute acceleration" (WAM 2025) | • Law No. 37 (2014) establishes CITRA with parliamentary oversight<br>• Digital authority split among Cabinet, Parliament, CITRA and CAIT | **High vs. low concentration of agenda-setting power**. UAE's concentrated authority accelerates policy adoption; Kuwait's overlapping jurisdictions create ≥ 4 veto points, slowing decisions |
| **Collective choice** (how decisions are made) | **Citizen-centricity**. Require citizen feedback panels or "happiness" KPIs in project design<br><br>**Collaborative governance**. Form interministerial AI steering committees and public–private working groups<br><br>**Public-value creation**. Align annual budgets and audit criteria with declared public value targets | • UAE Council for AI (Cabinet, 2018).<br>• UAE NAS 2031 (targets US$91 bn growth)<br>• Universal AI Blueprint 2024: every entity appoints a Chief AI Officer<br>• Dubai AI Policy 2025.<br>• All reference UAE PDPL 2021 and Cybercrime Law 34/2021, imposing cumulative fines AED 5 k – 5 m (US$1360–1.36 million) for missed API deadlines or data breaches | • Central Agency for Information Technology launches the Kuwait NAS (2024) (Central Agency for Information Technology 2024)<br>• Kuwait-DPPR covers only telecom & IT<br>• High-profile Google Cloud MoU (Jan 2023) still idle | **Single AI council vs. multiple committees; credible fines vs. soft exhortation**. UAE's enforceable sanctions and unified forum cut transaction costs; Kuwait's dispersed veto points and nonbinding guidance stall collaboration |
| **Operational** (day-to-day implementation) | **Citizen-centricity**. Implement AI-driven services that guarantee 24/7 assistance like chatbots, set response time benchmarks, and grievance channels<br><br>**Collaborative governance**. Define API standards and cost-sharing formulas for real-time data exchange.<br><br>**Public Value Creation**. Establish dashboards that track value outputs, legitimacy, and capacity | • Mandatory cross-agency data-sharing decree (2021)<br>• UAE-PDPL 45/2021 + Executive Regulations: data officer mandate, transparency duties, grievance timelines<br>• Circulars specify audit triggers and DPO reporting lines<br>• Dubai's AI Use-Case Scoring Matrix reallocates documented savings to new pilots (Digital Dubai and Dubai Future Foundation 2025) | • Sector-specific privacy regulation (DPPR 2024)<br>• Ministry circulars govern pilot approvals; no cross-agency grievance standard | **Operational clarity vs. uncertainty**. UAE's rule bundle fosters experimentation and reinvestment; Kuwait's compliance ambiguity deters pilots and prolongs "test before trust" cycles |

## 3 | Methodology

Grounded in Elinor Ostrom's IAD framework (Ostrom 2005, 58–59), this study traces how rule configurations in two Gulf states, the United Arab Emirates (UAE) and Kuwait, shape public sector deployment of AI. This section explains (i) the comparative logic and case selection criteria; (ii) the data-collection streams and processing pipeline; (iii) the analytical framework and coding steps; and (iv) the validity safeguards, ethics, and limitations. Figure 1 summarizes the sequential, multimethod design.

### 3.1 | Comparative Logic and Case .Selection

The study employs a combined Most Similar/Most Different Systems Design (MSSD/MDSD) to isolate institutional effects (Divald 2023). The UAE and Kuwait are alike in key structural drivers of digital government, hydrocarbon wealth, high broadband penetration, and small populations, yet they differ markedly in their constitutional and collective choice rules: the UAE is a federal system that shares power between the federal government and autonomous emirates, whereas Kuwait is a unitary state with a centralized government, an elected parliament, and a constitutional monarchy (Section 2.4; Mandelli 2025). By holding structural factors constant, we can observe how rule configurations shape AI's contribution to post-NPM goals.

To illustrate mechanisms inside each system, both country arms combine document analysis with semi-structured interviews. In the UAE, we examine two embedded cases, Smart Dubai and the Mohamed bin Zayed University of AI (MBZUAI) (Ministry of Human Resources and Emiratisation and Dubai Residency 2024; Mohamed bin Zayed University of Artificial Intelligence 2025). Smart Dubai is further unpacked into two subcases, the Dubai Happiness Meter and the DubaiNow app, which together embody all three post-NPM dimensions: citizen-centricity, collaborative governance, and public value creation (Smart Dubai 2021). MBZUAI foregrounds public value creation through talent development and research commercialization. These initiatives were selected because (a) both are government-initiated flagships launched for more than 18 months, (b) together they span the full post-NPM spectrum, and (c) extensive multiyear documentation and 19 relevant interviews are available. The Kuwaiti arm mirrors this strategy, relying on 20 interviews and the same document collection protocol; no single AI project in Kuwait has yet reached the level of maturity seen in the UAE. Therefore, this analysis focuses on the regulatory environment shaping emerging pilot initiatives. The Sahel app is a prominent example of a government-led digital platform in Kuwait. Operating for over 18 months and involving multiple agencies, its core functionality depends on APIs to integrate services from various ministries. However, there is no evidence that the Sahel app is actively incorporating machine learning (ML) or AI into the platform at this stage.

The study therefore asks:

**RQ**. How do constitutional, collective choice, and operational rules in use shape the effective adoption of AI in Kuwait and the UAE, and in turn advance post-NPM goals?

### 3.2 | Data Collection Streams

Three complementary evidence streams build a triangulated corpus (Table 2).

- Document processing pipeline

Official web domains, for example (ai.gov.ae, citra.gov.kw) were scraped with targeted keywords ("artificial intelligence,"

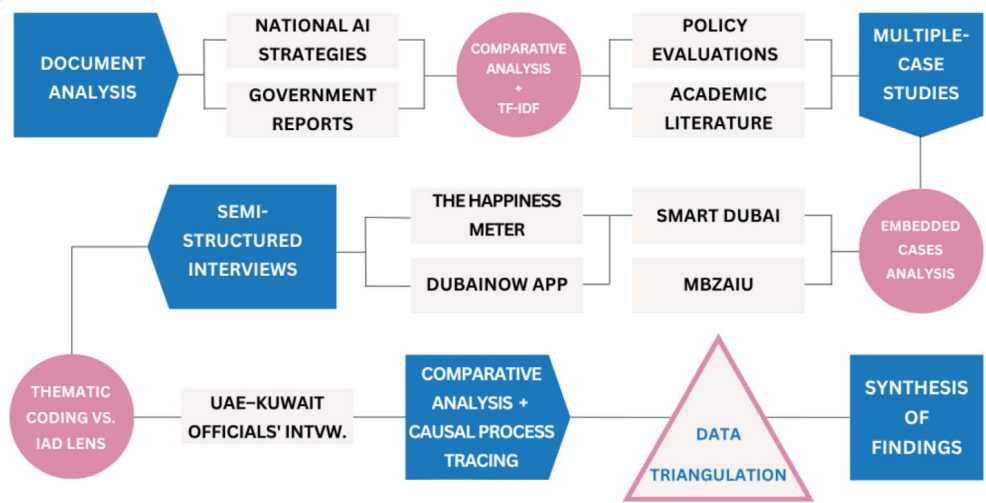

**FIGURE 1** | Overview of research design.

**TABLE 2** | Data sources and sampling parameters.

| Stream | Corpus/sample | Inclusion criteria & period |
| --- | --- | --- |
| Document corpus | 62 items, (UAE = 35; Kuwait = 27) documents totaling 1338 pages | Issued between February 2018 (when the UAE NAS appeared) and March 2025; formally endorsed by a federal ministry, royal decree, or parliament; explicitly reference AI, data governance, or digital government reform; and are publicly accessible or obtained through freedom of information (FOI) requests |
| Embedded cases | Smart Dubai & MBZUAI, internal evaluations, white papers, dashboards (14 docs, 638 pages) | Mandated by NASs; in operation ≥ 18 months; multiagency scope; employ ML components |
| Semi-structured interviews | $n = 39$, (Kuwait = 20; UAE = 19) | Adults with formal decision authority or advisory roles on AI. Kuwait sample spans senior civil servants, frontline managers, private developers, civil society advocates, academics, entrepreneurs. UAE sample comprises senior experts and scholars recruited via professional networks. Conducted Aug 2024—May 2025; thematic saturation at interview 18 (Kuwait) and 17 (UAE) (Fusch and Ness 2015) A deductive design seeks theoretical sufficiency rather than exhaustive theme discovery; we followed "Meaning saturation" to ensure that each prespecified construct is richly populated by participant experience (Hennink and Kaiser 2022, 7) |

"machine learning," "data protection") between February 2018 and March 2025. PDFs were converted to machine-readable text via *pdfplumber*; 16% of pages required OCR (Adobe + Tesseract) with an error rate < 5%. Arabic texts were professionally translated into English and back-translated. A bilingual glossary ensured terminological consistency.

– Interview protocol

Between August 2024 and May 2025, we conducted 39 semistructured interviews, meeting participants face to face or, when geography required, via encrypted video calls after recruitment through professional networks such as LinkedIn. A purposive "reputational" sampling strategy (Neuman 2014, 279) began with Ministry of Communications officials and snowballed outward. Each participant received an information sheet, signed informed consent in Arabic or English, and was assured anonymity (codes P1–P39). The 13-item guide probed rule clarity, formal and informal rules, data-sharing incentives, and perceived alignment with post-NPM goals. Sessions averaged 54 min (range 28–79), were audio recorded, transcribed verbatim, and anonymized excerpts were returned to a subset of participants for member checking within seven days ($n = 8$).

Conducting interviews in hierarchical Gulf public sector settings poses risks of perceived coercion and socially desirable responding. We therefore (i) recruited outside line management chains via professional networks; ministries were informed about the study in general terms but not about individual participation; (ii) conducted only one-to-one interviews, scheduled offsite or via encrypted video at times chosen by participants; (iii) emphasized voluntariness in the consent script, including the right to skip any question or stop at any time, and offered an "off the record" option for sensitive remarks; (iv) ensured interviewers had no supervisory, evaluative, or contractual ties to participants; (v) used neutral wording, counterfactual probes, and requests for negative cases to reduce social desirability bias; (vi) extended confidentiality beyond personal names by masking agency identifiers and using role-level descriptors such as "senior civil servant" or "municipal technology officer" where disclosure could risk identification; and (vii) kept reflexive interviewer memos after each session to note potential power asymmetries and adjust subsequent prompts. These steps complemented member checking and our broader triangulation across documents, embedded cases, and interviews.

### 3.3 | Analytical Framework and Coding Steps

We unify our definitional anchors, indicator set, and source triangulation plan for the three post-NPM goals (Table 3). The validity safeguards are summarized in Table 4. We then implement a three-stage coding pipeline in NVivo 14:

1. Open coding, text segments tagged to one of nine parent nodes (three rule tiers × three post-NPM goals).

2. Axial coding, relationships mapped, such as "constitutional rule → collaborative governance bottleneck."

3. Selective coding, cross-case mechanisms extracted, such as how ministerial vetoes discourage data pooling.

To make the coding logic visually transparent, Figure 2 maps the three dimensions of public value onto the Strategic Triangle that informs our analysis.

Two researchers independently coded 25% of the material; intercoder agreement was Krippendorff's $\alpha = 0.81$ (Krippendorff 2016). Discrepancies were resolved through memo-based consensus. Detailed procedures for unit-of-coding rules, multicoding, adjudication, and memoing are provided in the Section 3.6.

**TABLE 3** | Operationalization of the three post-NPM goals.

| Post-NPM goal | Definition | Comprehensive indicators | Source triangulation |
|---|---|---|---|
| Citizen-centricity | Placing citizens at the core of service design and delivery, emphasizing accessibility, responsiveness and personalization. In AI settings this means digital platforms that engage users directly, collect feedback and adapt services in real time<br>*Source:* Denhardt and Denhardt (2015); Kuziemski and Misuraca (2020); Osborne and Gaebler (1992) | • Accessibility of digital channels/channel reach<br>• Presence of real-time feedback or two-way loops<br>• Depth of personalization in service menus and notifications<br>• Transparency artifacts/visibility of algorithmic decisions<br>• Response-time benchmarks for chatbots & hotlines | Policy documents; service dashboards; interviews with managers & citizen advocates |
| Collaborative governance | Coordinated networks of government, private firms, civil society and citizens that coproduce policy and services, transcending siloed administration<br>*Source:* Ansell and Gash (2008); Ansell and Torfing (2021); Christensen and Lægreid (2013, 2022); Christensen (2012); Meijer and Bekkers (2015) | • Number and breadth of cross-agency data links/legally binding APIs<br>• Existence and authority of multi-actor steering bodies<br>• Formal partnerships or MoUs with private and civil society actors<br>• Joint budget or cost-sharing rules<br>• Citizen co-production mechanisms (crowdsourcing, hackathons) | -Statutes, MoUs, committee minutes, interviews with public private actors |
| Public value creation | The extent to which government action generates societal benefit beyond narrow efficiency metrics, i.e., outcomes valued by the community, sustained public trust and economic progress<br>*Source:* Bryson et al. (2014); Moore (1997); Weigl et al. (2024) | • Value outputs that users or society regard as beneficial, including labor hours saved and cost per transaction<br>• Legitimacy and support from authorizing environments (parliamentary endorsement, user-trust scores)<br>• Operational capacity: staffing, data, finance that sustain an initiative beyond pilot stage<br>• Appropriated vs. declared budgets; audit findings; initiative longevity | Budget papers; performance dashboards; external evaluations |

– **Explanation building**. We used causal process tracing to connect specific rule configurations to observable outcomes through identifiable mechanisms. Competing explanations, such as sheer fiscal capacity or vendor lobbying, were subjected to hoop tests, and we accepted a mechanism only when it was corroborated by at least three independent sources.

### 3.3.1 | TF–IDF Lexical Analysis

We use term frequency–inverse document frequency (TF–IDF) descriptively to gauge the relative rhetorical emphasis of concepts in strategy texts; it does not serve as a hypothesis test (Robertson 1977). Texts were lowercased and dehyphenated variants were harmonized, like (citizen-centricity ⇄ citizen centricity). Tokenization used whitespace/punctuation rules; numerals and punctuation were dropped. A standard English stop word list was applied; country and institution names were retained. We did not stem or lemmatize to preserve key phrases. To capture multiword concepts, we included uni, bi, and trigrams ($n = 1$–3) and report aggregated weights for concept variants, for instance (public value creation/public value). For each NAS, term frequency is sublinear $tf(t, d) = 1 + \ln f(t, d)$; inverse document frequency is computed across the full 62document corpus to stabilize weights, with smoothing as in Equation (1):

$$idf(t) = \ln\left(\frac{1+N}{1+df(t)}\right) + 1, \quad w(t, d) = tf(t, d) \times idf(t), \tag{1}$$

where $N$ is corpus size and $df(t)$ is document frequency. Document vectors are L2normalized. Reported scores are therefore comparable across the two NAS documents and reflect concept rarity in the wider corpus. All occurrences were included, whether in titles, informative statements, or normative statements (Lewis et al. 2020). Calculations were performed in Python 3.11 using scikit-learn (v1.4; TfidfVectorizer; Pedregosa et al. 2011) and NLTK (v3.8); see the Replication materials (Section 3.6) for the exact parameter settings.

### 3.4 | Validity Safeguards

To minimize bias and strengthen credibility, Table 4 summarizes the methodological safeguards employed.

*Politics & Policy*, 2025　　　　　　　　　　　　　　　　　　　　　　　　　　　　　　　　　　　　　　　　　　　　　9 of 20

**TABLE 4** | Validity threats and mitigation measures.

| Threat | Mitigation |
| --- | --- |
| Publication/selection bias (rules visibility) | Bilingual keyword scraping; snowballing from cited instruments; FOI requests where permissible; manual checks on ministry portals; versioning notes in corpus log; treat "rules on paper" as indicative |
| Rules on paper vs. enforcement in practice | Require ≥ 2 independent sources for enforcement claims (such as statute/circular + interview/dashboard); qualify sanction-related findings when verification is not possible |
| Translation & OCR noise | Professional translation and backtranslation; bilingual glossary; OCR validation with spot error < 5%; human review of high salience passages |
| Temporal drift/policy churn | Timestamp all coding (as of 1 May 2025); report the scrape window (2018–2025); avoid evergreen claims; flag areas most sensitive to later reforms |
| Case maturity asymmetry (UAE vs. Kuwait) | Focus Kuwait analysis on the regulatory environment and emerging pilots; avoid scale comparisons; emphasize mechanism identification over performance ranking |
| Coding reliability | Dual coding of 25% of materials; Krippendorff's $\alpha = 0.81$; memo-based adjudication |
| Interview reactivity | Triangulation with written sources; member checks with eight participants; neutral prompts and requests for negative cases |
| Power asymmetries & social desirability | Nonmanager-mediated recruitment; private/offsite or encrypted interviews; explicit voluntariness in consent; role-level masking; reflexive interviewer memos; triangulation |
| Researcher positionality | Mixed team composition (two GCC national coauthors; one nonregional auditor); external audit of coding memos |

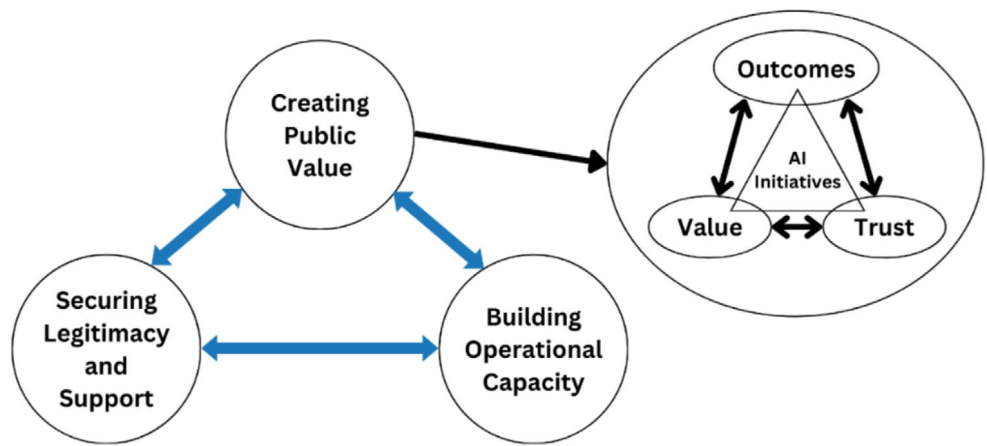

**FIGURE 2** | Strategic triangle framework, *Source:* Adapted from Moore (1997).

## 3.5 | Ethical Compliance

Fieldwork received approval from the AGU Kuwait IRB (Protocol #58924) and AGU UAE IRB (Protocol #32125). Participants were informed of their right to skip questions or withdraw; none exercised this option. Personal identifiers were removed from transcripts, and encrypted audio files were deleted after verification.

## 3.6 | Data & Materials Availability

All materials needed to evaluate and replicate the analysis are available in the Harvard Dataverse repository (DOI: 10.7910/DVN/NZKWS6). The repository includes: the coding protocol; the final codebook; an anonymized quote table; the exact TF–IDF configuration (software/versions/parameters); an NVivo node-tree outline; and the document-variable schema used for crosstabs. Raw interview recordings/transcripts are not shared.

## 3.7 | Limitations

This two-country design limits regional generalizability, but the depth of rule mapping supports theory-building inference. Publicly available documents may omit unpublished directives or rescinded circulars; we therefore treat "rules on paper" as indicative rather than exhaustive and qualify claims accordingly. Formal provisions (such as fines, deadlines) may also be unevenly enforced; we triangulate enforcement with interviews and public notices/dashboards and interpret sanction-related findings cautiously. Translation/OCR noise is possible (16% of pages required OCR); professional translation with back-translation, a bilingual glossary, and OCR spot checks (< 5%) were used to minimize error, which may affect lexical counts at the margin rather than qualitative themes. AI-related rules evolve rapidly; all coding is current as of May 1, 2025. Despite safeguards outlined in Section 3.2, interviews in politically sensitive, status-stratified settings can reflect organizational loyalties or social-desirability bias; we mitigated this via non-manager recruitment, private scheduling, neutral prompts, role-level masking, member checking, and triangulation, but residual bias may remain. Regarding TF–IDF, (i) it captures salience in wording, not importance or implementation; (ii) results are sensitive to authorial style; and (iii) multiword concepts can be underestimated if phrased idiosyncratically. To illustrate coding without breaching confidentiality, see the anonymized quote table in the (Section 3.6) provides short, role-masked excerpts with triangulation pointers; raw transcripts remain confidential under IRB protocols.

## 4 | Findings

Drawing on the triangulated evidence base described in Section 3, this chapter reconstructs how concrete rule configurations in the UAE and Kuwait translate or fail to translate into the three post-NPM outcomes of citizen-centricity, collaborative governance, and public value creation. Results are reported in four blocks that match the study's data streams:

1. Documents corpus—rules on paper
2. Embedded cases—rules in action
3. Semi-structured interviews—rules in formation
4. Cross-case process tracing—why similar resources yield different outcomes

### 4.1 | Documents Corpus: Rule Reach and Thematic Focus

Since the launch of the UAE NAS in 2018, the Emirates have issued 15 AI-specific documents and binding decrees, compared to just one from Kuwait. The UAE's regulatory output has grown steadily, culminating in a notable spike in 2024, while Kuwait began contributing only in that year. This 15:1 disparity underscores the UAE's sustained institutional commitment to AI governance, while Kuwait's entry signals a more tentative, delayed engagement. Figure 3 visualizes rulemaking intensity over time, a proxy for institutional commitment that we link to scaling outcomes in Section 4.4, and it illustrates this stark contrast. Counts reflect items meeting the inclusion criteria in Table 2.

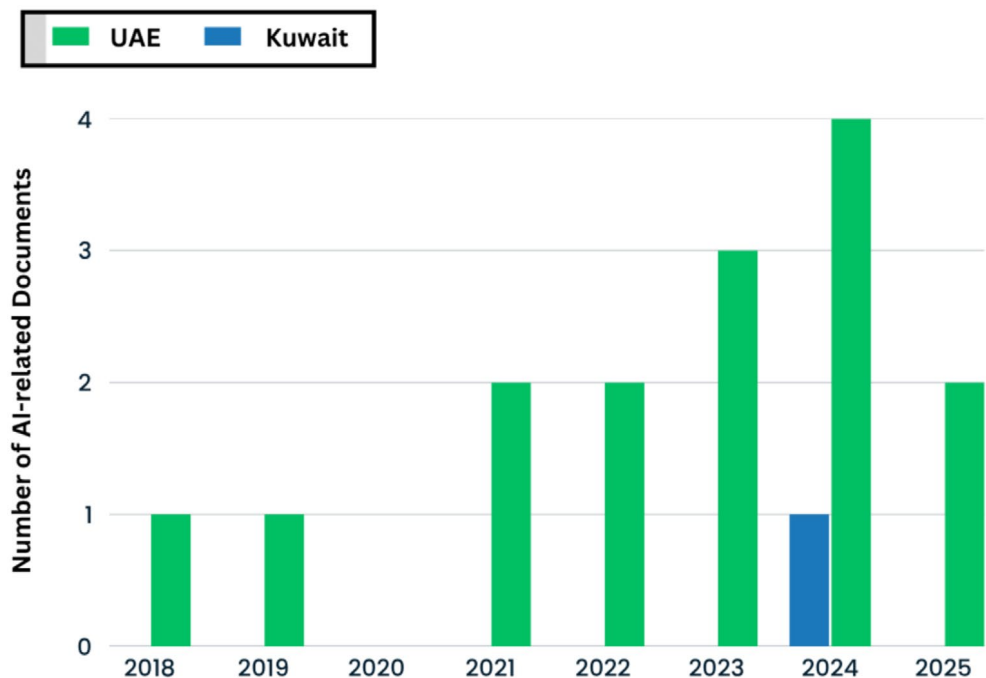

**FIGURE 3** | Annual production of AIR-related public documents in the UAE and Kuwait, 2018–2025.

The nature of these documents, however, is just as important as their number. Each policy tool can be categorized by its bindingness, how directive or enforceable it is. This distinction is essential: binding tools have a direct impact on implementation and compliance, shaping real-world AI deployment.

**4.1.1** | **Strategic Language and Topic Salience**

The two NASs texts distribute attention unevenly across post-NPM concepts. Table 5 lists the raw TF–IDF scores.

Figure 4 compares post-NPM thematic emphasis in the two NAS texts, helping assess whether rhetorical priorities align with observed implementation patterns. It translates those weights into a heat map, making each strategy's thematic "center of gravity" visually obvious.

Remarkably, the Kuwaiti NAS allocates its highest lexical weight to "collaborative governance," yet field evidence shows persistent operational gridlock. This mismatch suggests that rhetorical commitment alone does not overcome dispersed veto points; indeed, the more actors invoked in strategy texts, the greater the coordination burden in practice, precisely the mechanism our interviews exposed (P7 KW, P13 KW); conversely, the UAE text foregrounds public value language, consistent with the output-oriented metrics reported in the Smart Dubai and MBZUAI cases.

**4.1.2** | **Capability Roadmap**

Capability terms provide further nuance. Figure 5 shows the capability focus each NAS foregrounds, which we use to interpret capacity constraints and talent flows reported in interviews; it then ranks the most distinctive AI capabilities in each NAS, based on TF–IDF weights.

Both NASs prioritize AI governance and cybersecurity, acknowledging the global shortage of advanced technical skills. Kuwait makes only scant mention of deep learning, reflecting interviewee concerns about a "brain drain to Dubai." Since TF–IDF reflects perceived strategic weight, these textual signals must be interpreted within the context of each country's stage of AI maturity. TF–IDF, of course, is not a definitive measure of need; higher scores merely indicate greater rhetorical emphasis. Nonetheless, rhetorical focus often shapes budget allocations, which, in turn, influence capability acquisition.

**TABLE 5** | IAD lexical focus. TF–IDF weights for post-NPM concepts in the UAE and Kuwait NASs.

| Country | Year issued | Terms | Frequency score | Total score | Normalized TF | IDF | TF-IDF |
|---|---|---|---|---|---|---|---|
| UAE | 2018 | Citizen-centricity | 5 | 25 | 0.20 | 0.176 | 0.0352 |
| | | Collaborative governance | 9 | | 0.36 | 0.176 | 0.06336 |
| | | Public value creation | 11 | | 0.44 | 0.176 | 0.07744 |
| Kuwait | 2024 | Citizen-centricity | 4 | 15 | 0.267 | 0.176 | 0.04699 |
| | | Collaborative governance | 7 | | 0.467 | 0.176 | 0.08219 |
| | | Public value creation | 4 | | 0.267 | 0.176 | 0.04699 |

*Note:* TF uses sublinear scaling $1 + \ln f$; IDF computed over the 62-document corpus with smoothing; document vectors L2-normalized; n-grams (1–3) included; English stop-words removed; hyphenation variants aggregated. Scores indicate relative rhetorical emphasis, not policy priority or effect size.

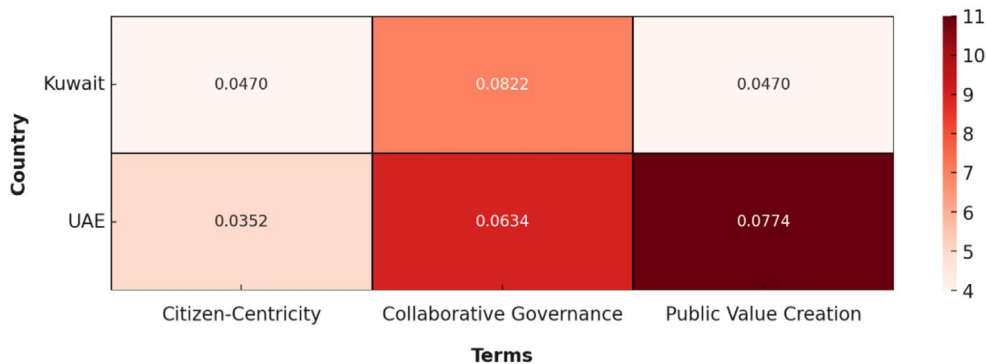

**FIGURE 4** | Heat map of normalized TF–IDF weights for selected post-NPM concepts in the two NAS documents. Values are comparable across NAS texts because IDF is estimated over the 62 document corpus; darker cells denote higher lexical salience. Descriptive only; sensitive to phrasing. See Table 5 and Section 3.3.1 for methodological details.

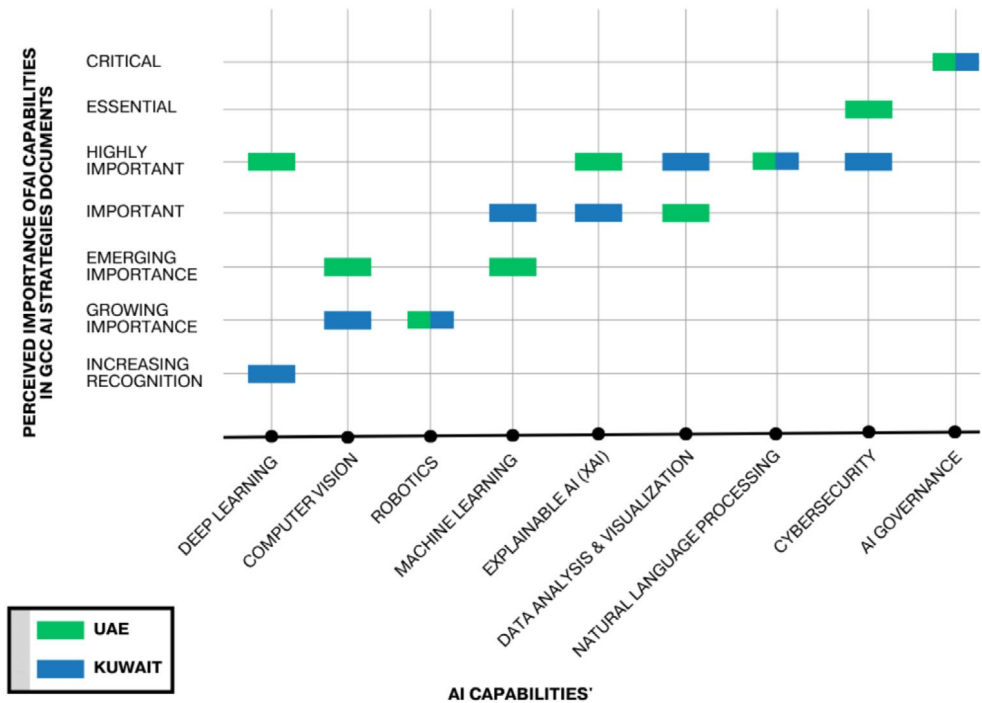

**FIGURE 5** | Relative salience of AI capability terms in the two NASs.

### 4.1.3 | Humancentric Rhetoric and AI Ethics

Phrases such as "give this human potential the best opportunities to nourish and flourish" (UAE NAS) and "invest in human capital development" (Kuwait NAS) map onto the legitimacy vertex of Moore's Strategic Triangle. AI ethics are mentioned extensively in both NASs, with repeated references to fairness, transparency, accountability, and privacy; and so on, the two strategies appear ethically adequate. Yet, each document devotes far more verbiage to economic growth and innovation than to safeguarding human agency, and crucially, neither prescribes an enforceable mechanism, such as a comprehensive AI law, mandatory impact audits, external oversight boards, or monetary sanctions, to make those ethics actionable. The result is a noteworthy gap between principled rhetoric and operational reality, leaving both countries vulnerable to the very "ethics-washing" dynamic discussed in Section 2.2.

## 4.2 | Embedded Cases: Rules in Action

### 4.2.1 | Smart Dubai (Royal Decree 2013)

**4.2.1.1 | The Happiness Meter.** Captures real-time sentiment and feedback from 4400 customer touchpoints across 192 entities. Since its launch, average satisfaction has risen from 90% in 2018 to 96% in 2024; the same dashboard reports a classic NPM efficiency gain: −time on task saved reduced 7.7 million labor hours and AED 725 million ($195.75 million USD) in staff cost avoidance. However, the sentiment analysis weighting methodology remains unclear. This AI-enabled system leverages sentiment analysis to rapidly identify service gaps and improvement areas based on user-generated inputs, from on-site kiosks to social media platforms. It automatically flags low satisfaction scores and triangulates textual feedback with other usage data, enabling AI models to continuously inform decision-makers about emerging issues or areas needing policy intervention. As a result, the initiative exemplifies *citizen-centric* service, allowing government agencies to proactively refine delivery and better respond to evolving public needs.

**4.2.1.2 | The DubaiNow App.** Consolidates over 280 services from 44 public and private sector entities, enabling everything from bill payments to license renewals and urban issue reporting (such as via Madinati). Since its launch in 2015, the platform has exemplified smart governance through seamless integration and high performance; 91% of API calls are processed in under 300 milliseconds. Agencies can be fined under UAE–PDPL regulations for delays in API integration, demonstrating the government's commitment to enforcement and accountability. For the flagship DomesticWorkers Bundle, the *unit cost per transaction* fell by AED 400 ($108 USD), and the *processing time* collapsed from ≈30 days to 5 days, tangible NPM cost-efficiency results inside a collaborative platform.

AI-powered modules underpin real-time data exchange and predictive analytics, ensuring smooth cross-agency coordination and effective partnerships with private-sector players including telecoms, banks, and municipal authorities. This high level of interoperability showcases AI's ability to manage vast datasets, flag inconsistencies, and automate backend processes that once demanded manual oversight. In doing so, DubaiNow exemplifies *collaborative governance*: AI-driven dashboards and decision-support systems reduce bureaucratic silos and streamline interdepartmental workflows.

However, achieving this technical integration required extensive stakeholder alignment and thoughtful AI system design,

*Politics & Policy*, 2025　　　　　　　　　　　　　　　　　　　　　　　　　　　　　　　　　　　　　　　　　　　　　　　　　　　　13 of 20

highlighting the complexity of deploying multi-stakeholder AI projects at scale in hybrid public–private ecosystems.

### 4.2.2 | MBZUAI (Federal Law 2019)

Heralded as the world's first graduate-level university dedicated exclusively to AI, MBZUAI hosts MSc and PhD students from 45 countries. Between 2019 and mid-2024, its academic community, comprising 80 faculty members, 200 researchers, and several hundred students, produced more than 300 peer-reviewed papers, secured 42 patents, and released Jais, the first open-source Arabic LLM. Ranked among the global top 20 by CSRankings in computer vision, NLP, and ML, with high *outsourcing intensity*, many active MoUs with private or foreign organizations channel research expenditure through external partners such as Google, Microsoft, IBM, Cerebras Systems, Presight AI, ADGM, Silal, and, since 2025, École Polytechnique, an NPM-style market mechanism that has deepened collaboration on joint initiatives to lower and streamline costs by reducing overlapping research budgets and unlocking new funding efficiencies. Its academic stature is further reinforced by the presence of distinguished visiting scholars like Michael Jordan and Sir Michael Brady. A state-of-the-art supercomputing cluster supports its research, which is integrated with ministerial sandboxes to address public sector challenges. This alignment between research and national priorities strengthens the UAE's AI capacity and drives meaningful *public value creation*.

## 4.3 | Semi-Structured Interviews: Rules in Formation

Thirty-nine semi-structured interviews, twenty in Kuwait, nineteen in the UAE, were coded against the IAD tiers. Consolidating both country streams here avoids scattering quotations and allows direct contrast.

### 4.3.1 | Constitutional Tier

Kuwait: Fourteen participants decried the absence of a single decision center. A senior Chief Information Officer (CIO) (P7 KW) remarked, "The Cabinet, CITRA and CAIT can all veto a dataset, three signatures, three calendars". UAE: An executive adviser (P2 AE) offered the counterpoint: "When the AI Office says 'connect', ministries ask 'by when', not 'why.'" Concentrated authority compresses lead times in Abu Dhabi; diffusion dilates them in Kuwait.

### 4.3.2 | Collective Choice Tier

Kuwait: An engineer at CAIT (P13 KW) observed that "Parliament wants fortnightly updates before we even test a chatbot," extending pilot phases well beyond regional norms. UAE: A municipal Chief Technology Officer (CTO) (P5 AE) noted that the UAE-PDPL and Cybercrime Law duo provides "a readymade template for any new MoU, legal review in a week."

Templates lower transaction costs and enable the 44 entity spread evident in DubaiNow.

### 4.3.3 | Operational Tier

Kuwait: Nine respondents pointed out that current public sector accounting rules classify expenditures on cloud-based graphics-processing units (GPUs), the hardware backbone for advanced data analytics and simulation projects, as long-term capital assets. Due to the restriction that unspent capital allocations cannot be reprogrammed once the fiscal year has commenced, several interviewees observed that a substantial portion of the 2024/25 digital-innovation budget remained unused. UAE: A program manager (P6 AE) offered a contrasting picture: "If my GPU rental overshoots, the Scoring Matrix lets me draw from savings another team banked." Flexible rebudgeting converts worldwide GPU scarcity into deliverable services. Two provosts at Kuwait University (P16 KW, P19 KW) warned that "shrinking domestic research funds, combined with the global techskills shortage, is fueling an accelerated 'braindrain to Dubai' trend." External benchmarks echo the interview narrative: the Stanford AI Index 2025 ranks the UAE #3 worldwide on net AI talent migration, while Kuwait is absent, consistent with respondents' "braindrain to Dubai" warnings.

### 4.3.4 | Public Trust Paradox

Kuwait sits first worldwide in the International Telecommunication Union (ITU) ICT Development Index 2023 and 2024, yet ministries fear privacy backlash. A tech scholar (P18 KW) summarized: "People do not trust the algorithm, the obstacle is policy (enforceable rights), not bandwidth or fiber-optic cables." UAE respondents describe the mirror dynamic: "Citizens assume UAE-PDPL protects them, so uptake is quick" (P3 AE).

### 4.3.5 | Foreign Cloud Deadlock

Several Kuwaiti officials cited Google Cloud's January 2023 framework agreement: since then, no datacenter license has cleared Cabinet and Parliament scrutiny. The stalemate is emblematic of overlapping mandates: CITRA, CAIT, the Cabinet, and the Parliament each reserve final say, creating an approval maze absent from the UAE.

## 4.4 | Cross-Case Process Tracing: Why Similar Resources Yield Different Outcomes

Synthesizing documents corpus, embedded cases, and thirty-nine semi-structured interviews yields a coherent causal story. In the UAE, decision rights are tightly concentrated: the Minister of State for AI, the federal AI Office, and, since 2025, the AI Legislative Intelligence Office sit at the apex of a single command chain. That concentration shortens feedback loops; agencies receive one set of marching orders and a clear deadline. Kuwait, by contrast, disperses authority among the Cabinet, CITRA, CAIT, and Parliament. Every major dataset or procurement must navigate four calendars and, frequently, four risk assessments, a sequence that interviewees themselves label "orchestrated gridlock."

The second differentiator is sanction credibility. Emirati interviewees speak of the UAE-PDPL (2021) and the Cybercrime Law (2021) in the same breath as project plans because both statutes threaten fines of AED 50,000 to 5 million for missing an API deadline or mishandling personal data. Penalties are automatic, codified in executive regulations, and enforced by a single inspectorate. Kuwaiti counterparts receive "letters of concern." With no monetary penalties, exhortations drift, and API integration stalls.

Narrative framing adds a third layer. Dubai brands itself "the world's happiest city," and federal projections promise US$91 billion in incremental AI-driven growth by 2031. Optimism legitimizes data sharing and helps citizens tolerate experimentation. Kuwait's storyline is the inverse, a "testbeforetrust" ethic, voiced explicitly by several senior officials, shapes decision cycles, and the Google Cloud framework agreement signed in January 2023 still awaits cross-agency clearance, symbolizing the perception that foreign clouds threaten data sovereignty.

Finally, resource discipline differs sharply. Under Dubai's AI Use Case Scoring Matrix, agencies must reinvest documented efficiency gains into subsequent pilots. Smart Dubai alone reports AED 725 million in direct savings and 7.7 million staff hours redeployed to frontline services; these funds constitute an internal venture capital pool for new AI applications. Kuwaiti budget rules take the opposite tack: cloud GPUs are booked as capital assets, and any unspent capital allocation lapses at the close of the fiscal year. Consequently, brain drain accounts for the UAE's net inflow of AI specialists and Kuwait's corresponding outflow, reinforcing the initial institutional asymmetry.

Taken together, these four mechanism bundles—rule concentration, sanction credibility, narrative alignment, and resource discipline—explain why the UAE scales proofs of concept into production services while Kuwait remains locked in pilot mode, despite comparable hydrocarbon revenues and a world-leading ICT infrastructure score.

## 5 | Discussion

This section interprets the findings through Ostrom's IAD lens and the hybrid NPM/post-NPM debate, draws design lessons, sets out case-specific implications for both the UAE and Kuwait, and then turns to theoretical/normative contributions, transferability, and a research agenda.

### 5.1 | Implications for Post-NPM Reform (Design Lessons)

Three design lessons emerge from the findings. First, concentrated constitutional authority can accelerate adoption, but only if paired with procedural transparency that guards against technocratic drift. Second, statutory sanctions at the collective choice tier are indispensable; collaboration flourishes when actors know that missed milestones carry real costs. Third, flexible operational budgets and clear grievance channels turn digital promise into citizen-centric value, reallocating underspent funds, publishing service-level objectives, and auditing algorithmic fairness at regular intervals.

### 5.2 | Case-Specific Policy Implications

#### 5.2.1 | United Arab Emirates: Preserve Speed, Strengthen Legitimacy

Because the UAE already exhibits high vertical rule coherence and credible sanctions, its binding risk is legitimacy and contestability, not delivery speed. The following oversight additions target that gap without diluting execution capacity.

- Constitutional (who may decide).
  - **(C1)** Establish by statute an independent Algorithmic Accountability and Audit Office with budgetary autonomy, mandated to issue an annual public report to the Federal National Council.
  - **(C2)** Codify a right to explanation and redress for consequential automated decision-making (ADM), with authority to compel remedial action in eligibility, licensing, or enforcement cases.
- Collective choice (How Decisions Are Made).
  - **(CC1)** Create a Public Register of High-Risk ADM Systems used by government, noting purpose, datasets, risk tier, and a point of contact; require a 30-day comment window before new high-risk deployments.
  - **(CC2)** Mandate Algorithmic Impact Assessments (AIAs) prelaunch and on major updates; publish nonsensitive AIA summaries.
  - **(CC3)** Require annual third-party audits (technical + process) for high-risk systems; include procurement clauses for model cards, data sheets, versioning, and log retention to ensure auditability.
  - **(CC4)** Earmark 5%–10% of documented efficiency savings to fund oversight capacity (audits, ombuds services) and small grants for civil society participation.
- Operational (daytoday implementation)
  - **(O1)** Standardize grievance channels with published SLAs, appeal ladders, and escalation to an independent digital rights ombuds; publish anonymized quarterly case statistics.
  - **(O2)** Require predeployment redteaming/bias testing with documented test suites; publish post-incident reviews for material failures "algorithmic incident reporting."
  - **(O3)** Publish public summaries of DPIAs (with necessary redactions) and maintain decision logs for the duration of appeal windows.
  - **(O4)** In sensitive domains, such as policing, sanctions, and eligibility, run time-boxed sandboxes with community panels and an option for human-only review during trials.

### 5.2.2 | Kuwait: Unlock Scale by Improving Rule Coherence and Sanction Credibility

Kuwait's binding constraint is fragmented authority, exhortative sanctions, and budget rigidities that keep initiatives in "pilot mode." Adding new oversight layers at this stage would likely create additional veto points. The emphasis should be on simplifying decision rights, enforcing timelines, and enabling reinvestment, with proportionate transparency measures.

- Constitutional (who may decide).
  - **(C1)** Designate by statute a single Government AI Coordinator with authority to set cross-agency integration deadlines and to arbitrate disputes on data exchange (clarifying the CAIT–CITRA boundary).
  - **(C2)** Require that AI-relevant executive circulars be consolidated and indexed centrally (official gazette or portal) to reduce ambiguity over rules in use.
- Collective choice (how decisions are made).
  - **(CC1)** Enact binding API integration timelines with graduated monetary sanctions for noncompliance; publish a simple MoU template pack that references privacy/security duties to cut legal cycle time.
  - **(CC2)** Create a "Lite" Register of government ADM use (purpose, legal basis, data categories) without prelaunch comment requirements, aimed at transparency without adding veto points.
  - **(CC3)** Standardize benefits realization reporting (time on task saved; cost per transaction) and link it to a reinvestment rule.
- Operational (day to day implementation).
  - **(O1)** Reclassify cloud GPUs and managed AI services as OPEX where appropriate and permit midyear virements to avoid lapsing funds; publish a short finance circular explaining treatment.
  - **(O2)** Adopt a reinvestment rule that recycles documented savings into the next cohort of pilots, with a rolling quarterly allocation.
  - **(O3)** Stand up a Shared API Catalogue with common schemas and a "ready-to-use" integration kit; publish grievance SLAs and a single cross-agency helpdesk to reduce fragmentation for citizens.
  - **(O4)** Use time-boxed sandboxes, such as 6–9-month trials, with predefined exit criteria (scale/stop/iterate) to prevent permanent pilots.

In short, UAE should add ex post transparency and independent scrutiny to an already coherent, sanction-backed rule stack; Kuwait should first simplify authority, make sanctions credible, and unlock budget flexibility, then layer in transparency and audit as scale is achieved.

### 5.3 | Theoretical and Normative Implications

The evidence refines Ostrom's core proposition by showing that alignment across IAD tiers (not strength at any single tier) predicts progression from pilot to production. It also supports a contingent hybridization view: NPM instruments (time on task, cost per transaction, outsourcing intensity) complement post-NPM aims only when embedded in enforceable rule bundles with reinvestment mechanisms. Narrative framing operates as a cultural-cognitive catalyst: pro-innovation narratives paired with visible safeguards lower perceived risk; risk-averse discourse without credible pathways raises transaction costs.

Normatively, a tension remains: can rapid, sanction-backed rollouts coexist with deep democratic deliberation? The UAE model excels at speed and scale but risks limited contestation; Kuwait safeguards pluralism at the cost of momentum. A tiered design (centralized authority for technical standards plus ex-post oversight, grievance channels, and public registers) offers a plausible middle path.

### 5.4 | Transferability and Research Agenda

The four mechanism bundles (rule concentration, sanction credibility, narrative alignment, resource discipline) form a portable diagnostic for other resource-rich settings. Where vertical coherence and enforcement are high, the next step is legitimacy safeguards; where authority is dispersed, the priority is coherence and credible deadlines, not more veto points. Future work should therefore: (i) assemble versioned, panel datasets of statutes, decrees, and enforcement events to track rule diffusion; (ii) link official narratives to behavior using computational linguistics; (iii) test blended scorecards that braid efficiency and legitimacy indicators; and (iv) couple mobility datasets with survey panels to trace how rule bundles shape AI talent flows.

### 6 | Conclusion

This study asked how constitutional, collective choice, and operational rules in use shape public sector AI adoption in two resource-rich GCC states. Using Ostrom's IAD lens across a triangulated corpus (documents, embedded cases, interviews), we identified four mechanism bundles (rule concentration, sanction credibility, narrative alignment, and resource discipline) that explain why the UAE converts pilots into operating services while Kuwait remains pilot-heavy despite comparable fiscal capacity. The core inference is simple: rules beat resources; it is vertical rule coherence, not wealth, that best predicts AI's public value yield.

### 6.1 | Theoretical Propositions

**P1**. Progression from pilot to production increases with alignment across IAD tiers; strength at any single tier is insufficient.

**P2**. Classic NPM instruments (time on task, cost per transaction, outsourcing intensity) complement post-NPM aims only when embedded in enforceable rule bundles with reinvestment mechanisms.

**P3**. Proinnovation framing paired with visible safeguards lowers perceived risk and facilitates collaboration; risk-averse framing without pathways to assurance raises transaction costs.

## 6.2 | Policy Implications

For centralized systems like the UAE, the binding risk is legitimacy and contestability, not delivery speed. A pragmatic path is to preserve rule concentration while adding ex-post transparency and independent scrutiny (public registers of high-risk ADM, algorithmic impact assessments and published summaries, annual third-party audits, standardized grievance SLAs with escalation, and earmarking a small share of efficiency gains to fund oversight and civic input).

For veto-laden systems like Kuwait, the binding constraint is fragmented authority and exhortative sanctions. The priority is to simplify decision rights, set and enforce integration deadlines with monetary sanctions, and unlock budget flexibility and reinvestment (OPEX-friendly treatment of cloud/AI services, rolling reinvestment of documented savings), complemented by a shared API catalogue and timeboxed sandboxes to avoid permanent pilots.

Our inference is bounded by reliance on publicly available documents (with possible undercoverage of informal rules), role-masked interviews, and a temporal cutoff of May 1, 2025; several outcome metrics are agency-reported and treated as indicative. These constraints are mitigated by triangulation and transparency of methods but should inform interpretation.

The four-mechanism diagnostic travels to other resource-rich polities: where coherence and enforcement are already high, add legitimacy safeguards; where authority is dispersed, first build coherence and credible deadlines, then layer oversight. Future work should track rule diffusion over time, test blended efficiency-legitimacy scorecards, and examine how narratives reshape citizen consent and talent flows.

In sum, without institutional coherence, even lavish AI budgets produce stalled pilots; with coherent, enforceable, and learning-oriented rule bundles, efficiency gains are more likely to convert into public value.

### Author Contributions

All authors contributed to the conceptualization and design of this study. Data collection and analysis were carried out by Mohammad Albous, while the interpretation of results was a collective effort by all authors. Mohammad Albous and Bedour Alboloushi prepared the original draft, and subsequent review and editing were undertaken by Bedour Alboloushi and Arnaud Lacheret. All authors provided final approval of the manuscript before submission.

### Acknowledgments

The authors gratefully acknowledge the invaluable support and insights provided by the government officials, policymakers, and academic experts in Kuwait who participated in this study. We also extend our thanks to the organizations and institutions that facilitated access to relevant data and resources. Finally, we appreciate the constructive feedback from the anonymous reviewers, which significantly contributed to improving the clarity and rigor of this manuscript.

### Conflicts of Interest

The authors declare no conflicts of interest.

### Data Availability Statement

The processed data supporting the findings of this study are available from the corresponding author upon reasonable request. Access may be subject to ethical or confidentiality constraints, depending on agreements with participants and data providers.